\documentclass[12pt]{article}
\usepackage{epsfig}

\def\Title#1{\begin{center} {\Large {\bf #1} } \end{center}}

\begin{document}

\Title{Radiative $B$ Decays  ---  an Experimental Overview}

\bigskip\bigskip


\begin{raggedright}  

{\it Edward H. Thorndike\index{Thorndike, E. H.}\\
University of Rochester\\
Rochester, NY 14627, USA\\
CLEO Collaboration}
\bigskip\bigskip
\end{raggedright}

\section{Introduction}

I'll give an informal, personal review of the status and direction of
experiments on radiative $B$ decays  ---  $b \rightarrow s \gamma$ and
$b \rightarrow d \gamma$.  Let's start by listing the observables.

\begin{itemize}

\item{the branching fractions for exclusive $b \rightarrow s \gamma$ decays, eg.
$B \rightarrow K^*(892) \gamma$}

\item{the branching fraction for the inclusive decay $b \rightarrow s \gamma$
(actually $B \rightarrow X_s \gamma$)}

\item{the $CP$ asymmetry in the inclusive decay and in exclusive decays: 
\newline $a_{CP} \equiv
 (\Gamma(b \rightarrow s \gamma) - \Gamma(\bar b \rightarrow \bar s \gamma))
/ (\Gamma(b \rightarrow s \gamma) + \Gamma(\bar b \rightarrow \bar s \gamma))$}

\item{the photon energy spectrum in inclusive decays $B \rightarrow X_s \gamma$}

\item{in principle, all the same observables for $b \rightarrow d \gamma$}

\end{itemize}

(In multibody final states, such as $B \rightarrow K \pi \pi \gamma$, there are
additional observables, constructed from the particle momenta.  I do not
consider these observables here.)

    What can each of these observables teach us?  The branching fractions for
exclusive $b \rightarrow s \gamma$ decays are the easiest of the observables,
and CLEO's observation~\cite{CLEO-K*-1} of $B \rightarrow K^*(892) \gamma$ back
in 1993 was the first penguin seen.  But while that exclusive decay was fine for
the `existence proof', the rates for exclusive decays are not useful for
searching for New Physics, because form factors are poorly known.

    In contrast, the branching fraction for the inclusive decay
$B \rightarrow X_s \gamma$ ($X_s$ a sum over all final states containing an $s$
quark), is ideal for revealing or limiting New Physics.  Forbidden at tree level
by GIM, the process proceeds via penguin diagrams.  In the Standard Model, the
loop contains $W^\pm$ and $t$, both heavy, and so New Physics penguins with,
eg., squarks and winos in the loop, would give comparable contributions.
Further, as a result of very hard theoretical work, the rate for
$b \rightarrow s \gamma$ can be reliably calculated, both within the Standard
Model and with New Physics.

    $CP$ asymmetries are very small in the Standard Model, 1\%
or less.  They can reach 10 - 20 \% in some New Physics proposals.  The
asymmetry for the inclusive process is more reliably calculated than that for an
exclusive process, but if a large $CP$ asymmetry is found in either, that will
be clear evidence of New Physics.

\begin{figure}[t]
\begin{center}
\epsfig{file=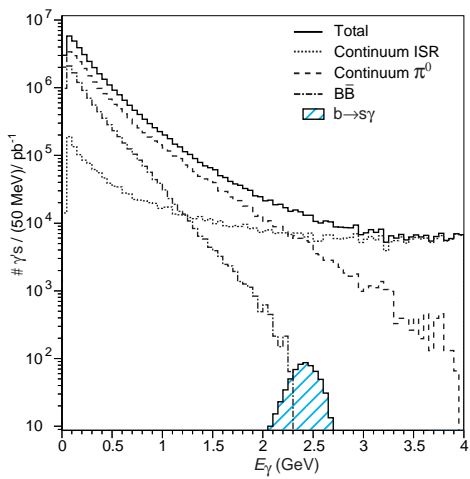,height=2.5in}
\caption{Photon energy spectrum expected from $b \rightarrow s \gamma$, other
$B$ decay processes, and from the continuum under the $\Upsilon(4S)$.}
\label{fig:glennplot}
\end{center}
\end{figure}

    In contrast to branching fractions, the photon energy spectrum in
$B \rightarrow X_s \gamma$ is very insensitive to New Physics.  The basic
process, $b \rightarrow s \gamma$, is a two-body decay, and hence gives a line
in the $b$ quark rest frame, broadened a bit by gluon bremsstrahlung.  The
photon energy spectrum for $B \rightarrow X_s \gamma$ thus depends on the mass
and Fermi momentum of the $b$ quark within the $B$ meson.  From the spectrum one
can learn the $B$ light cone shape function, useful for obtaining
$\vert V_{ub} \vert$ from the endpoint lepton yield in
$b \rightarrow u \ell \nu$.  Also, the spectrum helps determine HQET OPE
expansion parameters, needed for obtaining a precision value of
$\vert V_{cb} \vert$ from the $b \rightarrow c \ell \nu$ inclusive rate.

    The initial interest in $b \rightarrow d \gamma$ will be in determining
$\vert V_{td} \vert$ from the rates for exclusive decays
$B \rightarrow \rho \gamma$, $B \rightarrow \omega \gamma$.  But here one must
watch out for long distance effects and for additional CKM factors from c- and
u-quark loops.

    The experimental problems in studying radiative $B$ decays are illustrated
in Fig.~\ref{fig:glennplot}.  There one sees the photon energy spectrum
expected from radiative
$B$ decays, from other $B$ decay processes, and from the continuum under
$\Upsilon(4S)$ -- photons from initial state radiation and from decay of hadrons
(dominantly from $\pi^0 \rightarrow \gamma \gamma$).  While
$b \rightarrow s \gamma$ can be distinguished from other $B$ decay processes by
measuring the yield above 2.2 GeV, the contribution from the continuum -- two
orders of magnitude larger than the signal -- is a major challenge.  Techniques
for suppressing the continuum background are a MUST.  With such techniques, and
the power of full $B$ reconstruction, exclusive decay modes stand out above the
continuum background.  For the inclusive process $B \rightarrow X_s \gamma$, it
is essential to measure and subtract the continuum background, by running below
the $\Upsilon(4S)$.

    While $b \rightarrow s \gamma$ can be separated from other $B$ decay
processes by considering only the yield above 2.2 GeV, that approach is
inadequate for the precision of today's $b \rightarrow s \gamma$ inclusive
branching fraction measurements, and also for obtaining a useful photon
spectrum.  For these, one must go down to at least 2.0 GeV, understanding and
removing the substantial yield from other $B$ decay processes between 2.0 and
2.2 GeV.

    In subsequent sections I discuss branching fractions for exclusive
$b \rightarrow s \gamma$ decays; the branching fraction for the inclusive
$b \rightarrow s \gamma$ decay; $CP$ asymmetries; the photon energy spectrum;
$b \rightarrow d \gamma$ decays.  In Section~7, I summarize and give
conclusions.

\section{Branching Fractions for Exclusive $b \rightarrow s \gamma$ Decays}

    CLEO's 1993 observation~\cite{CLEO-K*-1} of eight
$B^0 \rightarrow K^{*0} \gamma$ events and five $B^+ \rightarrow K^{*+} \gamma$ 
events was based on 1.4 ${\rm fb}^{-1}$ of 4S luminosity.  With an order of
magnitude more luminosity, CLEO~\cite{CLEO-K*-2}, BaBar~\cite{BaBar-K*}, and
Belle~\cite{Belle-K*} now all have 10-20\% measurements of both charged and
neutral decays.  Results are given in Table~\ref{tab:K*br}.  Agreement among
measurements is good.  Branching fractions for charged and neutral decays
agree well.

\begin{table}[htb]
\begin{center}
\begin{tabular}{l|c|c}  
   & $B^0 \rightarrow K^{*0} \gamma$ &$B^+ \rightarrow K^{*+} \gamma$ \\ \hline
 CLEO '93\cite{CLEO-K*-1} & 4.0 $\pm$ 1.7 $\pm$ 0.8 & 5.7 $\pm$ 3.1 $\pm$ 1.1 \\
 CLEO '00\cite{CLEO-K*-2} & 4.55$\pm$0.70 $\pm$0.34 & 3.76$\pm$0.86$\pm$0.28 \\
BaBar '02\cite{BaBar-K*}  & 4.23$\pm$0.40 $\pm$0.22 & 3.83$\pm$0.62$\pm$0.22 \\
Belle(prelim)\cite{Belle-K*}&4.08$\pm$0.34 $\pm$0.26& 4.92$\pm$0.57$\pm$0.38 \\
 Average       & 4.21$\pm$0.25 $\pm$0.26 & 4.32$\pm$0.38$\pm$0.30 \\ \hline
\end{tabular}
\caption{$B \rightarrow K^* \gamma$ branching fractions ($10^{-5}$) }
\label{tab:K*br}
\end{center}
\end{table}

    In addition to $K^*(892)$, CLEO~\cite{CLEO-K*-2} and
Belle~\cite{Belle-new, Belle-K*}
have observed $B \rightarrow K^*_2(1430) \gamma$, with branching fractions
of $1.66 \pm 0.56 \pm 0.13 \times 10^{-5}$ and
$1.50 \pm 0.56 \pm 0.12 \times 10^{-5}$, in good agreement and of comparable
accuracy.  Belle has also~\cite{Belle-new} observed
$B^+ \rightarrow K^+ \pi^- \pi^+ \gamma$ ($2.4 \pm 0.5 \pm 0.3 \times 10^{-5}$),
and deduced substructures $B^+ \rightarrow K^{*0} \pi^+ \gamma$ ($2.0 \pm 0.65
\pm 0.2 \times 10^{-5}$), $B^+ \rightarrow K^+ \rho^0 \gamma$
($1.0 \pm 0.5 \pm 0.25 \times 10^{-5}$).  There is no evidence of a nonresonant
component, with upper limit ${\cal B}(B^+ \rightarrow K^+ \pi^- \pi^+ \gamma)NR
 < 0.9 \times 10^{-5}$.

\section{Branching Fraction for Inclusive $b \rightarrow s \gamma$ Decay}

    For a study of the inclusive process $B \rightarrow X_s \gamma$, lacking the
discrimination that comes from full $B$ reconstruction, continuum suppression is
very important.  In the `first observation of inclusive' analysis,
CLEO~\cite{CLEO-inc-1} used two approaches.  The first was to choose several
(eight) ``shape variables'', each with some power to discriminate between
$b \rightarrow s \gamma$ signal and continuum background (either ISR or
$\gamma$'s from hadrons), and combine them into a single variable using a neural
net.  (This was CLEO's first use of a neural net, and I was initially very
negative about the approach.  Its success made me a convert.)

    The second approach, dubbed ``pseudoreconstruction'', at first sight appears
just like full reconstruction.  Events containing a high energy photon are
searched for combinations of particles that satisfy $B \rightarrow X_s \gamma$. 
For $X_s$, we try one kaon ($K^\pm$ or $K^0_s \rightarrow \pi^+ \pi^-$), and 1
to 4 pions (of which at most one may be a $\pi^0$).  The measure of ``satisfying
$B \rightarrow X_s \gamma$'' is closeness of $M$ and $E$ to the proper values,
as given by
$\chi^2_B \equiv (E - E_{beam})^2/\sigma_E^2 + (M - M_B)^2/\sigma_M^2$.
A $\chi^2_B < 20$ is deemed an acceptible pseudoreconstruction.  What makes this
``pseudoreconstruction'', rather than full reconstruction, is our lack of
concern as to whether we ``have all the pieces right.''  True $B \rightarrow
X_s \gamma$ events are much more likely to pseudoreconstruct than are continuum
background events, and this remains true with one or two mis-chosen pions.
Further, $\cos \theta_{tt}$, the cosine of the angle between the thrust axis of
the particles that pseudoreconstruct and the thrust axis of the rest of the
event, is strongly peaked for the jet-like continuum events, but isotropic for
signal events.  (I was initially {\it very} dubious about this technique,
fearing that it would be very sensitive to the choice of model for
$B \rightarrow X_s \gamma$, but this proved not to be the case.)

    In CLEO's 1995 publication~\cite{CLEO-inc-1}, we performed two separate
analyses, one using shape variables with a neural net, the other using
pseudoreconstruction.  We then averaged the two branching fractions so obtained.
In CLEO's latest publication~\cite{CLEO-inc-2}, we did a fully integrated
analysis.  For all events with a high energy photon, we obtained a combined
shape variable parameter from the neural net (8 inputs, 1 output).  For the
subset of events that had a pseudoreconstruction with $\chi^2_B < 20$, we
obtained two additional discriminating parameters, $\chi^2_B$ and
$\vert \cos \theta_{tt} \vert$.  For the subset that contained a lepton ($e$ or
$\mu$), we used the energy of the lepton and the angle between lepton and high
energy photon as additional discriminating parameters.  Armed with these
discriminating parameters (sometimes only 1, sometimes 3, sometimes 5), we
determined the probability that an event with a high energy photon was
$b \rightarrow s \gamma$ rather than continuum background, and assigned it a
weight according to that probability.

\begin{figure}[htb]
\begin{center}
\epsfig{file=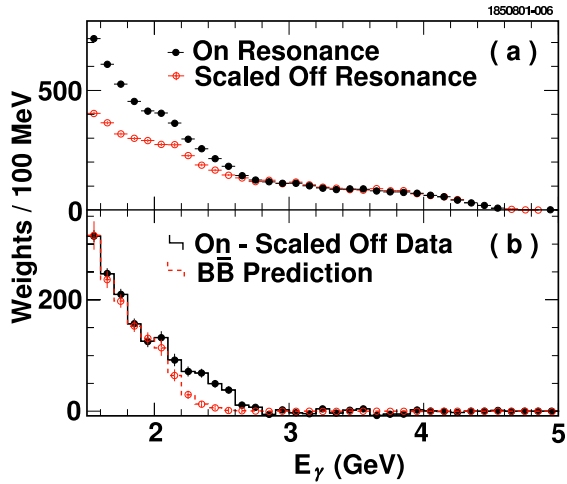,height=2.5in}
\caption{CLEO's~\cite{CLEO-inc-2} photon energy spectrum:  (a) On resonance and
scaled Off resonance; (b) On minus scaled Off, and prediction for $B \bar B$
processes other than $b \rightarrow s \gamma$ and $b \rightarrow d \gamma$.}
\label{fig:onoffspec}
\end{center}
\end{figure}

    The distribution in weights so obtained {\it vs.} photon energy is shown in
Fig.~\ref{fig:onoffspec}.  The upper panel shows On-resonance and scaled 
Off-resonance data.  The success of the continuum suppression is apparent, in
that the continuum background is now a mere factor of 4 larger than the signal,
rather than the two orders of magnitude in Fig.~\ref{fig:glennplot}.

    The lower panel in Fig.~\ref{fig:onoffspec} shows the yield in weights
{\it vs.} photon
energy after the continuum background has been subtracted, using Off-resonance
data.  There one sees clear evidence for $b \rightarrow s \gamma$ in the 2.2 -
2.6 GeV range, and also the increasing importance of the other $B$ decay
processes below 2.2 GeV.

    In its first measurement~\cite{CLEO-inc-1}, CLEO placed a cut on photon
energy at 2.2 GeV, using theoretical models to account for the fraction of
$b \rightarrow s \gamma$ rate below 2.2 GeV, and accepting a systematic error
for this model dependence.  In the recent measurement~\cite{CLEO-inc-2}, by
making a strenuous effort to understand background from $B$ decay processes,
CLEO lowered its photon energy cut to 2.0 GeV, thereby accepting $\sim$90\% of
the rate, and reducing the systematic error from model dependence.  To be
competitive, future measurements will have to accept photons down to at least
2.0 GeV.

\begin{figure}[htb]
\begin{center}
\epsfig{file=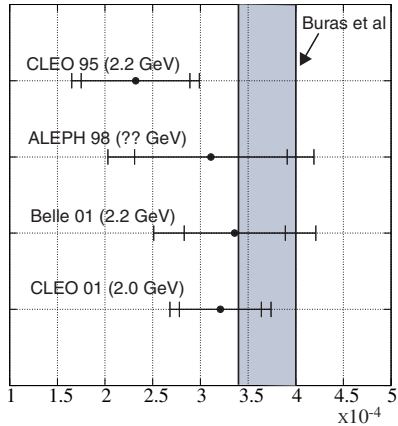,height=2.5in}
\caption{Measurements of the branching fraction for the inclusive process
$b \rightarrow s \gamma$, by CLEO in 1995~\cite{CLEO-inc-1}, ALEPH~\cite{ALEPH},
Belle~\cite{Belle-inc}, and CLEO in 2001~\cite{CLEO-inc-2}.  The Standard Model
prediction of Buras {\it et al.}~\cite{Buras} is also shown .}
\label{fig:brinc}
\end{center}
\end{figure}

    In addition to CLEO's two published measurements, there have been
measurements by ALEPH~\cite{ALEPH} and Belle~\cite{Belle-inc}. All four results
are shown in Fig.~\ref{fig:brinc}.  Difficult as the measurement is at the
$\Upsilon(4S)$,
it seems to me to be near impossible at the $Z^0$, and ALEPH's efforts must be
characterized as heroic. Their result is consistent with the
$\Upsilon(4S)$ measurements, but their error is twice that of the recent CLEO
measurement.  The Belle measurement, based on only 6 ${\rm fb}^{-1}$, and with
the now no-longer-acceptible 2.2 GeV photon energy cut, should be viewed as a
warmup exercise.

    The Standard Model theoretical expectation, as given most recently by
Buras {\it et al.} ~\cite{Buras}, is also shown in Fig.~\ref{fig:brinc}.
It should be mentioned that there have recently been questions
raised~\cite{Misiak} as to the appropriate value of $m_c/m_b$ to use in the
calculation.  Buras {\it et al.} have used
$m_c/m_b = m_c^{\overline{MS}}(\mu)/m_b^{1S} = 0.22$, while earlier work used
$m_c/m_b = m_c^{pole}/m_b^{pole} = 0.29$, and obtained a branching fraction
lower by $0.25 \times 10^{-4}$.  My impression is that the theoretical community
is not of a single mind as to the appropriate value of $m_c/m_b$ to use, and so
the SM theory value might come down by 10\%.  In any case, experiment and SM
theory are in fine agreement.

\section{$CP$ Asymmetries}

    There have been measurements of the $CP$ asymmetry in
$B \rightarrow K^*(892) \gamma$ by CLEO~\cite{CLEO-K*-2}, BaBar~\cite{BaBar-K*},
and Belle~\cite{Belle-K*}, and a measurement of the $CP$ asymmetry in
$B \rightarrow X_s \gamma$ plus $B \rightarrow X_d \gamma$ inclusive by
CLEO~\cite{CLEO-CP}. The {\it sign convention}, to my knowledge so far used in
all measurements of $B$ $CP$ asymmetries, is $b$ quark minus $\bar b$ quark
($B^-$ minus $B^+$, $\bar B^0$ minus $B^0$).  For the cases at hand, that means
$a_{CP} \equiv
(\Gamma(b \rightarrow s \gamma) - \Gamma(\bar b \rightarrow \bar s \gamma)/
(\Gamma(b \rightarrow s \gamma) + \Gamma(\bar b \rightarrow \bar s \gamma)$, and
similarly with the exclusive decays.

    Results for the exclusive decays are shown in Table~\ref{tab:K*aCP}.
CLEO's inclusive result, a combination of $b \rightarrow s \gamma$ and 
$b \rightarrow d \gamma$,
is $(0.965 a(b \rightarrow s \gamma) + 0.02 a(b \rightarrow d \gamma)) =
-0.079 \pm 0.108 \pm 0.022$.  All these results, including the average of the
three exclusive measurements, are consistent with zero, the Standard Model
expectation.

\begin{table}[htb]
\begin{center}
\begin{tabular}{l|c}  
   & $a_{CP}(B \rightarrow K^* \gamma)$  \\ \hline
 CLEO '00\cite{CLEO-K*-2}    &  +0.08  $\pm$ 0.13  $\pm$ 0.03  \\
BaBar '02\cite{BaBar-K*}     & --0.044 $\pm$ 0.076 $\pm$ 0.012  \\
Belle(prelim)\cite{Belle-K*} &  +0.032 $\pm$ 0.069 $\pm$ 0.020  \\
 Average                     &  +0.009 $\pm$ 0.048 $\pm$ 0.018  \\ \hline
\end{tabular}
\caption{$B \rightarrow K^* \gamma$ CP asymmetries}
\label{tab:K*aCP}
\end{center}
\end{table}

\section{The Photon Energy Spectrum}

    As can be seen from Fig.~\ref{fig:onoffspec}, in order to obtain the 
photon energy spectrum
for the inclusive $B \rightarrow X_s \gamma$ process, to photon energies of 2.0
GeV and below, one must understand backgrounds from $B$ decay processes.  Unlike
those from continuum processes, these cannot be directly measured.
CLEO~\cite{CLEO-inc-2} has proceeded as follows.

    The dominant component of the background, accounting for 90\%, is photons
from $\pi^0 \rightarrow \gamma \gamma$ and $\eta \rightarrow \gamma \gamma$ that
have escaped the $\pi^0$ and $\eta$ vetoes.  These backgrounds are determined by
measuring $\pi^0$ ($\eta$) yields, treating the $\pi^0$ ($\eta$) as if it were a
$\gamma$, applying all cuts and determining the event weight, just as in the
$b \rightarrow s \gamma$ analysis.  Monte Carlo is then used to determine the
$\pi^0$ ($\eta$) veto inefficiency.

    Photons from other sources are small by comparison to those from $\pi^0$ and
$\eta$, and with modest efforts to have the Monte Carlo event generator
accurate, one can (CLEO does) trust the Monte Carlo.  Processes considered
include $\omega \rightarrow \pi^0 \gamma$,
$\eta^\prime \rightarrow \rho^0 \gamma$, radiative $\psi$ decay,
$\rho \rightarrow \pi \gamma$, $a_1 \rightarrow \pi \gamma$, final state
radiation.  In addition to the dominant $b \rightarrow c$ decays,
$b \rightarrow u$ processes and $b \rightarrow s g$ processes were considered.

    Neutral hadrons, in particular antineutrons and K-longs, by interacting in
the calorimeter, cause high energy clusters, above 1.5 GeV.  Their contribution
to the $B$ decay background was determined by fitting the lateral distribution
of the shower (E9/E25, for those familiar with this notation).

\begin{figure}[htb]
\begin{center}
\epsfig{file=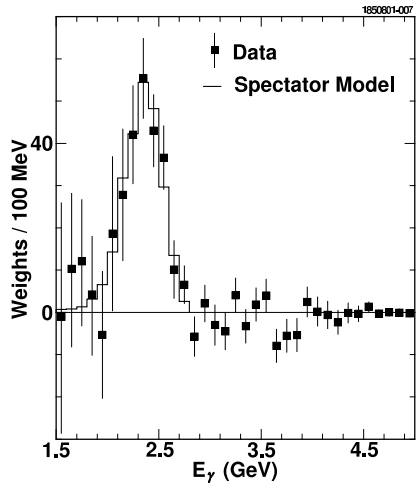,height=2.5in}
\caption{CLEO's~\cite{CLEO-inc-2} observed laboratory frame photon energy
spectrum for On minus scaled Off minus $B$ backgrounds, the putative
$b \rightarrow s \gamma$ plus $b \rightarrow d \gamma$ signal.}
\label{fig:spectrum}
\end{center}
\end{figure}

    CLEO's observed laboratory frame photon energy spectrum for On-resonance
minus scaled Off-resonance minus $B$ backgrounds (the $b \rightarrow s \gamma$
plus $b \rightarrow d \gamma$ signal) is shown in Fig.~\ref{fig:spectrum}.

    From the measured spectrum CLEO has obtained first and second moments, in
the $B$ rest frame, for $E_\gamma^{rest frame} > 2.0$ GeV, finding

$\langle E_\gamma \rangle = 2.346 \pm 0.032 \pm 0.011$ GeV, and

$\langle (E_\gamma - \langle E_\gamma \rangle )^2\rangle = 0.0226 \pm 0.0066
\pm 0.0020\ \ {\rm GeV}^2$.

\noindent HQET plus OPE allows inclusive observables to be written as double
expansions, in powers of $\alpha_s$ and $1/M_B$.  The parameter $\bar \Lambda$
enters at order $1/M_B$, $\lambda_1$ and $\lambda_2$ enter at $1/M_B^2$, and six
more parameters, $\rho_1$, $\rho_2$ and ${\cal T}_1 - {\cal T}_4$ at $1/M_B^3$. 
Using expressions in the $\overline {MS}$ renormalization scheme, to order
$1/M_B^3$ and order $\alpha_s^2 \beta_0$, CLEO obtains
$\bar \Lambda = 0.35 \pm 0.08 \pm 0.10$ GeV from the first moment.  The
expression for the second moment converges slowly in $1/M_B$, and CLEO did
not extract parameters from it.

    To lowest order in $\Lambda_{QCD}/M_B$, the hadron level
$B \rightarrow X_s \gamma$ photon energy spectrum is given by a convolution of
the parton level $b \rightarrow s \gamma$ photon energy spectrum with the
$b \rightarrow light quark$ light cone shape function of the $B$
meson~\cite{Kagan-Neubert}.  Again to lowest order in $\Lambda_{QCD}/M_B$, the
same shape function describes $B \rightarrow X_u \ell \nu$, i.e., the hadron
level $B \rightarrow X_u \ell \nu$ lepton energy spectrum is given by a
convolution of the parton level $b \rightarrow u \ell \nu$ lepton energy
spectrum with the same shape function~\cite{DeFazio-Neubert}.  CLEO has thus
used their measured $B \rightarrow X_s \gamma$ photon energy spectrum to
determine (to some accuracy) the light cone shape function, and from this
predicted the fraction of the $B \rightarrow X_u \ell \nu$ lepton energy
spectrum that lies above some cut near the endpoint.  This, combined with a
measurement of the $B \rightarrow X_u \ell \nu$ yield above that cut gives the
total $B \rightarrow X_u \ell \nu$ yield, and that in turn gives
$\vert V_{ub} \vert$~\cite{CLEO-endpoint}. Corrections enter at next order in
$\Lambda_{QCD}/M_B$, and these are currently the subject of active
investigation~\cite{Bauer, Ligeti, Neubert}.

\section{$b \rightarrow d \gamma$}

    So far there is nothing on inclusive $b \rightarrow d \gamma$.  On exclusive
$b \rightarrow d \gamma$, there are upper limits on
$B^+ \rightarrow \rho^+ \gamma$, $B^0 \rightarrow \rho^0 \gamma$, and
$B^0 \rightarrow \omega \gamma$.  From isospin and SU(3) considerations, one
expects ${\cal B}(B^+ \rightarrow \rho^+ \gamma) =
2 \times {\cal B}(B^0 \rightarrow \rho^0 \gamma) = 
2 \times {\cal B}(B^0 \rightarrow \omega \gamma)$.  Upper limits, from
CLEO~\cite{CLEO-K*-2}, Belle~\cite{Belle-BCP4}, and BaBar~\cite{BaBar-rho}, are
given in Table~\ref{tab:rholim}.  The BaBar limit is by far the best. Since
CLEO's first
observation of $B \rightarrow K^*(892) \gamma$ was based on 1.4 ${\rm fb}^{-1}$,
and since $B^+ \rightarrow \rho^+ \gamma$ is expected to be 20 times smaller
than $B \rightarrow K^* \gamma$, one can anticipate an observation by BaBar
and/or Belle in the near future.

\begin{table}[htb]
\begin{center}
\begin{tabular}{l|c|c|c|c}  
    & $B$ pairs & \multicolumn{3}{c}{Branching Fraction Upper Limits
($10^{-6}$)} \\ \cline{3-5}
   & (Million) & ${\cal B}(B^+ \rightarrow \rho^+ \gamma)$ &
$ 2 \times {\cal B}( B^0 \rightarrow \rho^0 \gamma)$ &
$ 2 \times {\cal B}( B^0 \rightarrow \omega \gamma)$ \\ \hline
 CLEO '00\cite{CLEO-K*-2}     & 9.7 & 13  & 34  & 18  \\
Belle '01\cite{Belle-BCP4}    & 11  & 10  & 21  & --- \\
BaBar(prelim)\cite{BaBar-rho} & 63  & 2.8 & 3.0 & --- \\ \hline
\end{tabular}
\caption{Upper limits on $B^+ \rightarrow \rho^+ \gamma$,
$B^0 \rightarrow \rho^0 \gamma$, and
$B^0 \rightarrow \omega \gamma$ branching fractions }
\label{tab:rholim}
\end{center}
\end{table}

    From their limit, BaBar~\cite{BaBar-rho} obtains
$[(1 - \rho)^2 + \eta^2]^{1/2}  < 1.6$.  This limit, while not an improvement in
the limit on $\vert V_{td} \vert$ over that obtained from the limit on
$B_s - \bar B_s$ mixing, provides nice confirmation.  But, I should repeat the
warning that, as accuracy improves, one needs to watch out for long distance
effects, and for contributions from c- and u-quark loops, carrying other CKM
factors.

\section{Summary and Conclusions}

{\bf $\mathbf{b \rightarrow s \gamma}$ Exclusive Branching Fractions}

\noindent These are no longer of great fundamental interest.  However, by
identifying a larger fraction of the makeup of $B \rightarrow X_s \gamma$
decays, one will reduce some systematic errors on the branching fraction for the
inclusive process $b \rightarrow s \gamma$.  Belle has made progress on this
front.  Perhaps more important, their observation of
$B \rightarrow K \pi \pi \gamma$ lays the groundwork for looking at correlations
among the momentum vectors of the decay products, providing a way to ``measure''
the helicity of the photon.

{\bf $\mathbf{b \rightarrow s \gamma}$ Inclusive Branching Fraction}

\noindent Experiment agrees well with the predictions of the Standard Model, and
places strong restrictions on New Physics.  But there is really only one good
measurement, CLEO's.  BaBar and Belle need to get to work on this one.  They
will need to accept photons down to 2.0 GeV or lower -- 2.2 GeV is no longer
good enough.  They will also need to take sufficient data at beam energies below
the $\Upsilon(4S)$, as the continuum subtraction {\it must} be done with
{\it data}.

{\bf CP Asymmetries}

\noindent So far there is no hint of a non-zero value.  Present limits place
weak restrictions on some New Physics models.  There is {\it plenty} of room for
improvements, with BaBar and Belle's large data samples, before systematic
error limitations set in.  Asymmetry measurements for the {\it inclusive} decay 
are desirable (BaBar, Belle?).

{\bf Photon Energy Spectrum}

\noindent CLEO's photon energy spectrum has helped provide a precise
determination of $\vert V_{cb} \vert$ from the inclusive semileptonic decay
branching fraction, and (more important) a good determination of
$\vert V_{ub} \vert$ from the lepton endpoint yield in
$b \rightarrow u \ell \nu$, with {\it quantifiable errors}.  Measurements of the
spectrum will be key for future determinations of $\vert V_{ub} \vert$ from
inclusive $b \rightarrow u \ell \nu$.  Improved measurements of the spectrum are
highly desirable.

{\bf $\mathbf{b \rightarrow d \gamma}$ Searches}

\noindent So far there is nothing on inclusives, and only upper limits on
exclusives.  These limits are not yet an improvement in the limit on
$\vert V_{td} \vert $ over that provided by the limit on $B_s - \bar B_s$
mixing.  But with data samples of 100 ${\rm fb}^{-1}$, BaBar and Belle should
see $B \rightarrow \rho \gamma$.  Stay tuned.
\vspace{1cm}

    I have benefitted from interactions with my many CLEO collegues.  Particular
thanks are due to Dan Cronin-Hennessy for his assistance in preparing this talk
and writeup.

\end{document}